%% file: main_article.tex
\documentclass[final,1p,times]{elsarticle}

\usepackage{lineno,hyperref}
\usepackage{caption}
\usepackage{subcaption}
\usepackage{amsmath}
\usepackage{float}
\usepackage{hyperref}
\modulolinenumbers[5]

\journal{Journal of Computational Materials Science}

\begin{document}

\begin{frontmatter}
\title{A molecular dynamics investigation of the dependence of mechanical properties of steel nanowires on C concentration.}

\author{J.K. Liyanage}
\author{M.D.Nadeesha Tharundi}
\author{Laalitha S. I. Liyanage}
\address{Faculty of Computing and Technology,University of Kelaniya,Dalugama, Sri Lanka}
\cortext[correpondingAuthor]{Corresponding author}
\ead{laalitha@kln.ac.lk}
%% or include affiliations in footnotes:

\begin{abstract}
\input{Sections/abstract}
\end{abstract}

\end{frontmatter}

%\linenumbers
\section{Introduction}
\input{Sections/introduction}

\section{Methods}
\input{Sections/methods}

\section{Results}
\input{Sections/results}

\section{Conclusion}
\input{Sections/Conclusion}

\bibliographystyle{unsrt}
\bibliography{mybibfile}       

\end{document}

%% file: Sections/abstract.tex
The temperature dependence of mechanical properties of steel nanowires with varying carbon content were studied using molecular dynamics simulations. Four interatomic potentials were assessed, with the Modified Embedded Atom Method (MEAM) potential developed by Liyanage et al. \cite{Liyanage2014_Cementite} selected for its accuracy in predicting the behavior of BCC Fe, FeC in the B$_1$ rock salt structure, and BCC iron with carbon. Uniaxial tensile tests were conducted on FeC nanowires with carbon concentrations of 0-10 \% at temperatures ranging from 0.1 K to 900 K. Stress-strain curves were analyzed to determine Young's modulus, yield stress, and ultimate tensile strength (UTS). Results showed that Young's modulus increased between 0.1 K and 300 K but decreased between 600 K and 900 K with increasing carbon content. Both yield stress and UTS decreased progressively with higher carbon percentages. Common Neighbor Analysis revealed rapid formation of slip planes as carbon content increased and greater slip plane propagation at elevated temperatures, contributing to reduced nanowire strength. These findings provide insights into the influence of carbon content and temperature on the mechanical behavior of steel nanowires, which may inform the design of nanostructured steel materials for various applications.

%% file: Sections/introduction.tex
Nanomaterials have garnered significant attention in recent years due to their unique properties and potential applications across various fields. Among these, nanowires have emerged as particularly promising structures, offering exceptional characteristics that make them suitable for a wide range of research areas. Iron (Fe), a versatile element with numerous applications, has been extensively studied in its nanowire form. Moreover, the addition of carbon to iron has been shown to enhance the metal's strength, making iron-carbon (Fe-C) nanowires a subject of great interest for their mechanical properties and potential as advanced nanomaterials.\\
The effect of alloying elements, particularly carbon, on the properties of bulk iron has been well-documented. In general, the addition of carbon to bulk iron increases its strength. Ferrite, a polycrystalline structure consisting of iron and a small amount of carbon (0.0069 wt.\% C), exhibits higher strength compared to pure iron. Specifically, the strength of bulk ferrite is 20\% higher than that of iron in the [1 1 1] and [1 $\overline{1}$ 0] crystallographic directions. This increased strength is attributed to the interaction between carbon atoms and dislocations, which causes maximum stress. The magnitude of this strengthening effect depends on the positions of both carbon atoms and slip planes \cite{IZQUIERDOSANCHEZ2020100822}.\\
However, when transitioning to the nanoscale, the behavior of iron and iron-carbon structures changes significantly. Interestingly, both ferrite and iron nanowhiskers demonstrate considerably lower strength compared to their bulk counterparts. This phenomenon is largely due to the high surface-to-volume ratio of nanowhiskers, which obscures the strengthening effect of carbon atoms, particularly in square-section whiskers \cite{IZQUIERDOSANCHEZ2020100822}.\\
To better understand the behavior of iron and iron-carbon nanowires, researchers have turned to molecular dynamics (MD) simulations. These simulations provide valuable insights into the mechanical properties and deformation mechanisms of nanoscale materials. However, the accuracy and reliability of these simulations heavily depend on the choice of interatomic potentials.\\
%The selection of an appropriate interatomic potential is crucial for accurately predicting the mechanical properties of materials in MD simulations. For body-centered cubic (BCC) systems like iron, this choice becomes even more critical due to the complex nature of defect structures, slip systems, and twinning behavior in these materials. Embedded Atom Method (EAM) potentials, while reliable for many face-centered cubic (FCC) systems, may be less accurate for BCC systems. This discrepancy arises primarily from the non-planar core of screw dislocations in BCC systems, which leads to more complex slip behavior compared to FCC systems.
For BCC iron nanowires, Mendelev's EAM model \cite{Mendelev2003} has been found to be particularly accurate in reproducing results\cite{Sainath2017}. This potential has successfully predicted the twinning mode of deformations in $<100>$ BCC Fe nanowires \cite{sainath2016orientation}. Other potential models, such as Tersoff and ReaxFF, which are bond-order based, have shown better accuracy than EAM/MEAM potentials for predicting the elastic properties of iron \cite{morrissey2019atomistic}. Additionally, the Finnis-Sinclair embedded atom interatomic potential has demonstrated good agreement with experimental results for cohesive energy and bulk modulus calculations \cite{qiao2016molecular}.\\
 A significant challenge in simulating Fe-C system is that existing interatomic potentials for the iron-carbon system often suffer from qualitative flaws in describing even the simplest defects. This limitation makes the results of molecular dynamics simulations in more complex environments potentially unreliable \cite{hepburn2008metallic}. A well designed interatomic potential is critical for obtaining realistic results. For instance, Angel et al. \cite{IZQUIERDOSANCHEZ2020100822} reported that the tensile strength of both ferrite and iron nanowhiskers was considerably lower compared to bulk specimens. Similarly, Henriksson et al. \cite{henriksson2013atomistic} found that nanowires with high impurity concentrations (Cr, C) fracture at shorter elongations than pure Fe.\\
To address the limitations of existing potentials, researchers have developed new approaches. Hepburn et al. \cite{hepburn2008metallic} presented an empirical potential based on density functional theory insights, which correctly describes the interaction of carbon and iron across a wide range of defect environments. This EAM-form potential is suitable for billion-atom molecular dynamics simulations and has been used to study carbon interactions with dislocations and calculate elastic constants of pure iron.\\
Henriksson et al. \cite{henriksson2013atomistic} developed potentials that accurately reproduce lattice parameters, formation energies, and elastic properties of principal Fe and Cr carbides. These potentials also predict good results for energy curves of mixing Fe-Cr, formation energies of simple carbon point defects in Fe and Cr, and martensite lattice anisotropy. Lee et al. \cite{lee2006modified} reproduced fundamental physical properties of relevant elements and alloys, including elastic, structural, defect, surface, and thermal properties. This potential is particularly useful for investigating the interaction between carbon interstitial solute atoms and other defects such as vacancies, dislocations, and grain boundaries in iron.\\
Rajabpour et al. \cite{rajabpour2015calculating} compared various potentials for calculating the bulk modulus of iron and steel. They found that while the Lennard-Jones potential was inadequate, EAM, MEAM, and Tersoff potentials showed good agreement with experimental data for both iron and steel. Liyanage et al. \cite{Liyanage2014_Cementite} studied the structural, elastic, and thermal properties of cementite using a MEAM potential for Fe-C alloys. Their results showed good agreement with density functional theory and experimental values, accurately predicting properties of most crystals.\\
While significant progress has been made in simulating iron and iron-carbon nanowires, there remains a need for further research. Limited studies have been conducted on the simulation of tensile strength and mechanical properties of pure iron in both BCC and FCC structures, with even fewer focusing on Fe-C nanowhiskers. The mechanical properties employed in most MD simulations of Fe-C nanowhiskers appear to be reasonably low, and a fully reliable interatomic potential for examining the influence of Fe-C composition on tensile behavior is still lacking.\\
Recent studies have also employed machine learning techniques in combination with molecular dynamics simulations to accelerate the prediction of elastic properties of Fe-C systems, highlighting the potential of data-driven approaches in complementing conventional simulations \cite{Risalma17030601}. Temperature plays a crucial role in determining the mechanical properties of iron nanowires. Studies have shown that Young's modulus and yield strength generally have an inverse relationship with temperature. As temperature increases, a greater number of atoms gain sufficient energy to overcome energy barriers, leading to decreased mechanical properties \cite{li2017molecular}. Higher temperatures also affect the evolution of dislocations. For instance, in Fe nanowires subjected to torsion, no dislocation lines were observed at 62 ps at 20 K, whereas numerous dislocation lines formed at the same time point at 100 K \cite{qiao2016molecular}.\\
This temperature-dependent behavior has been observed in both bulk iron and iron nanopillars, with Young's modulus and yield strength decreasing as temperature increases \cite{kotrechko2009temperature},\cite{wang2014martensitic}. The thermal vibrations induced by increased temperature can lead to the nucleation of defects in crystals.\\ %Moreover, temperature affects the atomic velocity and kinetic energy of the atoms \cite{li2017molecular}.
Interestingly, BCC Fe nanowires undergo a ductile-to-brittle transition at around 400 K. Below this temperature, nanowires yield through the nucleation of sharp cracks and fail in a brittle manner, while at higher temperatures, significant plastic deformation occurs\cite{Sainath2017}.\\
Given the exceptional properties exhibited by materials at the nanoscale and the known strengthening effect of carbon on iron, Fe-C nanomaterials are expected to demonstrate remarkable mechanical properties. Therefore, it is crucial to conduct comprehensive simulations of the mechanical properties of Fe-C alloy nanowires in future research. Based on the literature review, four potentials have been identified as suitable for investigating the effect of carbon percentage and temperature on steel nanowires: two MEAM potentials \cite{lee2006modified},\cite{Liyanage2014_Cementite}, one Tersoff potential \cite{henriksson2013atomistic}, and the Hepburn EAM potential \cite{hepburn2008metallic}. These potentials, obtained from the National Institute of Standards and Testing (NIST) interatomic potential repository \cite{haleinteratomic}, will be used to evaluate and determine the most suitable potential for future studies on Fe-C nanowire systems.

%% file: Sections/methods.tex
Current study was commenced by evaluating four interatomic potential parameterizations for the Fe and FeC systems. Initial evaluation of interatomic potentials was done by calculating bulk properties. Bulk properties of BCC Fe, hypothetical $B_1$ rocksalt structure and BCC iron with 1\% – 20 \% of C atoms were calculated and compared with experimental or first principal data. The published interatomic potentials were obtained from the National Institute of Standards and Testing (NIST) interatomic potential repository \cite{haleinteratomic}.

\subsection{Evaluation of interatomic potential}
\subsubsection{Bulk properties of iron}
Bulk properties of BCC Fe were calculated using all four interatomic potentials \cite{Liyanage2014_Cementite}, \cite{hepburn2008metallic}, \cite{henriksson2013atomistic}, \cite{lee2006modified}. The results are given in Table \ref{tab:2}. Bulk properties such as bulk modulus, equilibrium lattice parameter, cohesive energy, and atomic volume were extracted from the energy-volume curves. The Birch-Murnaghan equation of state (EOS) \cite{murnaghan1944compressibility}, shown in equation \ref{eq:1}, was fitted to the energy-volume curves to extract the structural properties of equilibrium BCC Fe predicted by each interatomic potential. The Birch-Murnaghan equation of state is given in equation \ref{eq:1}, where \( E \) is the cohesive energy at a given volume \( V \), \( E_0 \) is the cohesive energy at equilibrium volume \( V_0 \), \( B_0 \) is the bulk modulus, and \( B_0^{\prime} \) is the derivative of the bulk modulus.

\begin{equation}\label{eq:1}
E(V) = E(V_0) + \frac{B_0 V}{B_0' (B_0' - 1)} \left[ B_0' \left( 1 - \frac{V_0}{V} \right) + \left( \frac{V_0}{V} \right)^{B_0'} - 1 \right]
\end{equation}

The energy-volume curve was plotted with a primitive cubic unit cell with the size of \(a_0 \times (1 \times 1 \times 1)\) containing 2 atoms. The equilibrium lattice constant \(a_0\) for BCC Fe is 2.86~Å \cite{lee2012atomistic}. Periodic boundary conditions were applied to all the x, y, and z directions. The energy minimization was performed at each volume according to the conjugate gradient method with a tolerance of \(1 \times 10^{-6}\) eV. The lattice constant was obtained by varying the lattice parameter in the range of 2~Å to 10~Å in 0.1~Å steps, and then from 2.8~Å to 2.9~Å in 0.01~Å steps.

\subsubsection{\texorpdfstring{Bulk properties of \(B_1\) crystal structure (rocksalt)}{Bulk properties of B1 crystal structure (rocksalt)}}

\begin{figure}[!ht]
    \centering
    \includegraphics[width=0.5\textwidth]{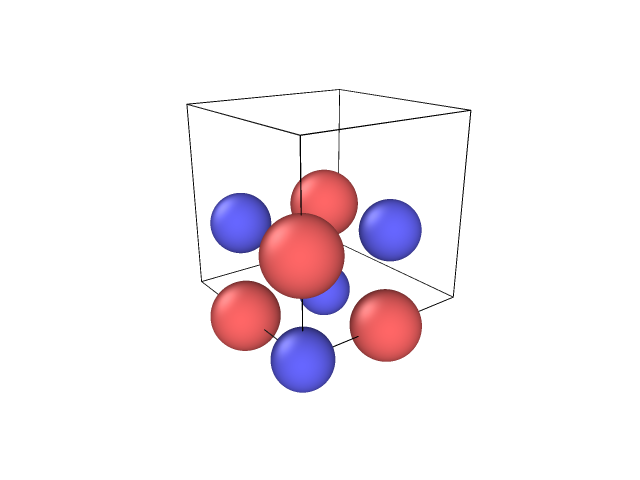}
    \caption{Unit cell \(B_1\) structure}
    \label{fig:unit cell B1 structure}
\end{figure}

Bulk properties of the hypothetical \(B_1\) structure were predicted using all the potential models. This structure contained the same concentration of 50\% of Fe and C each.
The energy-volume curve was plotted with a primitive unit cell containing 8 atoms with the size of \(a_0\) [\(1 \times 1 \times 1\)] in all 3 crystallographic directions. Figure \ref{fig:unit cell B1 structure} shows the unit cell of the \(B_1\) crystal structure. Periodic boundary conditions were applied to all the x, y, and z directions. At each volume, energy minimization was done with an energy tolerance of \(1 \times 10^{-25}\) eV.

The equilibrium lattice constant for the $B_1$ structure was 3.996 Å \cite{henriksson2013atomistic}. Bulk properties of $B_1$ crystal structure were extracted from the energy-volume curves. Results shown in Table \ref{tab:3}. The values predicted by DFT were used to set bulk properties, since experimental values were not available in this hypothetical structure. Predicted properties from interatomic potentials were compared with first-principles calculations using density functional theory (DFT). DFT was a first principles calculation, atomic interactions were not required to be modeled. It was much more accurate than MD and computationally much more expensive.

\subsubsection{Bulk properties of BCC Fe with C}
\begin{figure}[!ht]
    \centering
    \includegraphics[width=0.5\linewidth]{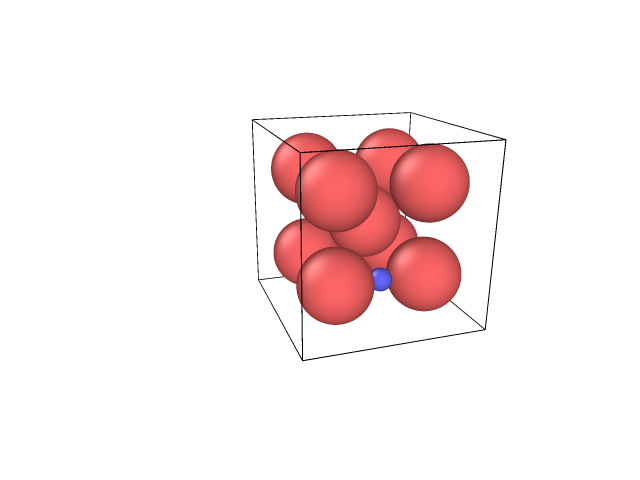}
    \caption{Unit cell of BCC iron with C interstitial atom at octahedral position}
    \label{fig:Unit cell of BCC iron with C interstitial atom at octahedral position}
\end{figure}

Effects of C on bulk structural properties were investigated through simulating BCC Fe with C at percentages ranging from 1\% - 20\%. Carbon was randomly distributed in octahedral sites of the BCC Fe structure. 
The energy-volume curve was plotted with a primitive unit cell of size \(a_0 \times [3 \times 3 \times 3]\) in all three crystallographic directions. Periodic boundary conditions were applied to all the \(x\), \(y\), and \(z\) directions. At each volume, energy minimization was performed with a tolerance of \(1 \times 10^{-25}\) eV. Figure \ref{fig:Unit cell of BCC iron with C interstitial atom at octahedral position} shows the three-dimensional view of the unit cell.

Due to the unavailability of experimental data for FeC system DFT calculations were performed for BCC Fe with 10 \% and 20 \% of Carbon. All C concentrations are in atomic percentage (at. \%). Weight percentages were not feasible to study because of the large simulation box size required to accommodate the required number of interstitial sites.
\subsection{Uniaxial tensile of steel NW}
\sloppy
Depending on the accuracy of the predictions, an interatomic potential was selected to model the mechanical properties of steel NWs. The nanowire simulation model was constructed in BCC structure with [100], [010], and [001] crystallographic directions in the x, y, and z directions. The Fe-C nanowire has a length of 121 Å in the z direction and a square cross-section with lengths of approximately 20 Å in the x and y directions. Substitutional octahedral C atoms were randomly distributed under the restriction that they are about one lattice parameter away from the boundary surface. Therefore, the probability of forming a graphite structure is zero.\\
The MD method was carried out in the LAMMPS package \cite{plimpton1995fast} to investigate the tensile behavior of the Fe nanowire and measure the temperature dependence due to various C percentages. Carbon concentrations of 0\%, 1\%, 5\%, 7\%, 9\%, and 10\% were simulated at temperatures of 0.1 K, 300 K, 600 K, and 900 K. Periodic boundary conditions were applied in all three directions. Before equilibration, the energy of the system was minimized using the conjugate gradient method with an energy tolerance of \(1 \times 10^{-25}\) eV. The simulation box was relaxed to optimize the volume according to the energy of the system, which iteratively changes the box size and the atomic coordinates until a minimum energy configuration is found. Then, the structure was equilibrated at the required temperature while keeping the number of atoms \(N\), pressure \(P\), and temperature \(T\) constant according to the NPT ensemble. The equilibration was run for 40 ps using a time step of 0.001 ps to bring the system to a stable temperature.

\begin{figure}[!h]
    \centering
    \includegraphics[width=0.5\linewidth]{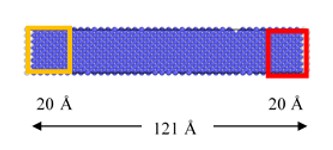}
    \caption{MD model for the Uniaxial Tensile Test}
    \label{fig:MD model for the Uniaxial Tensile Test}
\end{figure}

After that, the boundary in the z direction was changed into a shrink-wrap boundary . This was to give the structure a finite length in the z direction. Three regions were specified in the equilibrated nanowire Figure \ref{fig:MD model for the Uniaxial Tensile Test}. The bottom 1/6 of nanowire was kept fixed while the top 1/6 of the nanowire was moved by 0.05 Å, per each loading step. Tensile test was performed at a constant strain rate of \(1 \times 10^{8} \, \text{S}^{-1}\) in uniaxial \(z\) direction. Deformation was run for 5,000,000 time steps. These steps were repeated until the NW fractured. After optimization of the structure a velocity profile was applied to the atoms to bring the system to a certain temperature. Then a MD run of 20 ps was done with NVT ensemble at the specific temperature to equilibrate the system.

%% file: Sections/results.tex
\subsection{Validation of Interatomic Potentials}
\subsubsection{Bulk Iron (BCC Fe)} 

\begin{figure}[H]
    \centering
    \includegraphics[width=0.5\linewidth]{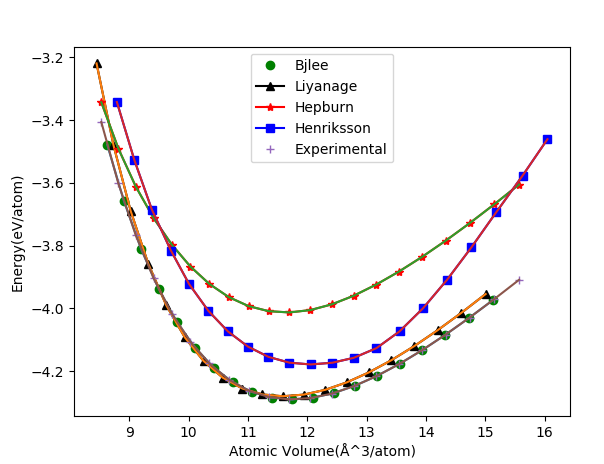}
    \caption{Energy vs Volume curves of Fe in BCC crystal structure predicted using the selected interatomic potentials and the experimental data.}
    \label{fig:Energy vs Volume curves of Fe in BCC crystal structure}
\end{figure}

Energy variation with respect to volume is considered an important test of validity for interatomic potential. The structural properties of BCC Fe are first investigated. Figure \ref{fig:Energy vs Volume curves of Fe in BCC crystal structure} demonstrates the energy-volume response of bulk Fe in comparison with curves generated by experimental data. The experimental curve is generated through the Birch-Murnaghan equation of state \cite{murnaghan1944compressibility} equation\ref{eq:1} using the experimental lattice constant \cite{lee2012atomistic}. 

\subsubsection{Energy-Volume curve}

EV curves generated using the potentials by B.J. Lee, Liyanage, Henriksson and Hepburn for pure BCC Fe are presented in Figure \ref{fig:Energy vs Volume curves of Fe in BCC crystal structure}. Bulk modulus (B), cohesive energy ($E_0$), equilibrium atomic volume (Ω$_0$) and equilibrium lattice constant are obtained by fitting the EV curves to the Birch-Murnaghan equation of state \cite{murnaghan1944compressibility} equation\ref{eq:1}. Compared to the structural properties calculated from the EV curves with experimental values the MEAM potential by Liyanage et al. \cite{Liyanage2014_Cementite}. and B.J. Lee et al. \cite{lee2006modified} . exhibit good agreement. The EAM potential \cite{hepburn2008metallic} is predicted to have a lower cohesive energy per atom and a lower bulk modulus for BCC Fe. Whereas the potential by Henriksson \cite{henriksson2013atomistic} predicts a higher bulk modulus with respect to other potentials. So, from the first test of evaluating interatomic potential parameterizations, MEAM potentials \cite{lee2006modified},\cite{Liyanage2014_Cementite} prove that they are suitable to predict the properties of the FeC system well with respect to other potentials. Structural properties compared to experimental values are given in Table \ref{tab:1}.

\begin{table*}[!h]
    \centering
    \caption{Comparison of the potentials by Liyanage, B.J. Lee, Henriksson, and Hepburn with experimental data for bulk Fe. $E_c$ is the cohesive energy (eV/atom), $a_0$ is the lattice constant (Å), B is the bulk modulus (GPa), and $\Omega_0$ is the atomic volume (Å\textsuperscript{3}/atom).}
    \label{tab:1}
    \begin{tabular}{lccccc}
        \hline
        Property & Expt\textsuperscript{\cite{lee2012atomistic}} & Liyanage & B.J. Lee & Henriksson & Hepburn \\
        \hline
        $E_c$ (eV/atom)       & 4.28     & 4.28   & 4.29    & 4.19       & 4.00 \\
        $a_0$ (Å)             & 2.86     & 2.85   & 2.86    & 2.89       & 2.86 \\
        $B$ (GPa)             & 166--173 & 179.08 & 166.33  & 229.78     & 153.66 \\
        $\Omega_0$ (Å\textsuperscript{3}/atom) & 11.70    & 11.58  & 11.73   & 11.96      & 11.64 \\
        \hline
    \end{tabular}
\end{table*}

\subsubsection{\texorpdfstring{FeC in NaCl rocksalt (B$_1$) structure}{FeC in NaCl rocksalt (B1) structure}}

Figure \ref{fig:unit cell B1 structure} shows the unit cell of the $B_1$ structure which is used in the current study to plot the EV curve of the $B_1$ rock salt structure which contains 8 atoms. (visualization of atomic configurations are performed using the OVITO package to obtain a unit cell of NaCl rocksalt structure.) Since Sodium chloride is a hypothetical structure, experimental data is not available. Therefore, predicted properties from interatomic potentials are compared with first principles calculations using DFT calculations.

\begin{figure}[!ht]
    \centering
    \includegraphics[width=0.5\linewidth]{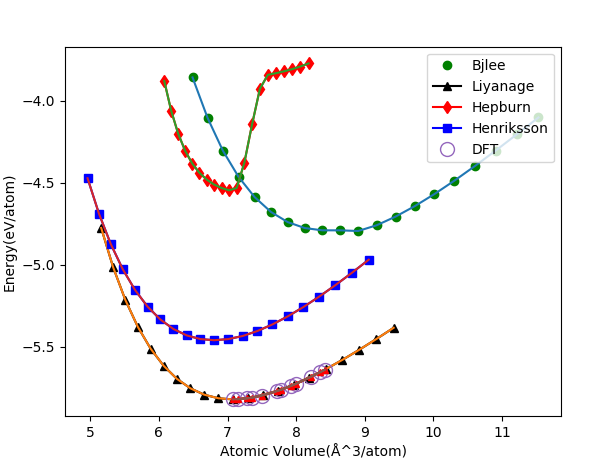}
    \caption{EvsV curves of $B_1$ rock salt structure}
    \label{fig:EvsV curves of B1 rock salt structure}
\end{figure}

\subsubsection{DFT curves} 

Comparison of EV curves with DFT for $B_1$ structure is presented in Figure \ref{fig:EvsV curves of B1 rock salt structure}. The EV curve of Liyanage MEAM potential clearly overlaps with the DFT curve. Therefore, it is evident that the Liyanage MEAM potential predicts the interactions between Fe and C with similar accuracy to the first principles DFT method. In comparison to the EV curves of BCC Fe, Liyanage MEAM has good agreement with experimental values. Hence, it can be concluded from the results gained so far that Liyanage MEAM potential is suitable for the simulation of both pure Fe and FeC systems. The structural properties such as cohesive energy, lattice constant, bulk modulus and atomic volume for the $B_1$ rocksalt structure are presented in Table \ref{tab:2}. 

In comparison with ab initio data the selected interatomic potentials by Liyanage, B.J. Lee, Henriksson and Hepburn perform at varying performances. The lattice constant of Liyanage potential exhibits good agreement with ab initio compared to other potentials. Also, the cohesive energy and atomic volume for the Liyanage potential predicts closer results to DFT values but the bulk modulus ~ 0.23\% less than DFT. The bulk modulus of Henriksson potential predicts closer results to DFT values. Contrary to this, Hepburn potential predicts the bulk modulus ~ 35\% greater than DFT. As mentioned earlier FeC $B_1$ structure is a hypothetical structure consisting of 50 \% Fe and 50 \% C atoms. Hepburn potential was developed to describe the interactions of carbon and iron under defect environments. Therefore, it isn't suitable for evaluating the structural or mechanical properties of carbon and iron in bulk structure. Hence, Hepburn potential does not perform well for high C concentrations. 

\begin{table*}[h]
    \centering
    \caption{$B_1$ structure properties by interatomic potentials B.J. Lee, Liyanage, Hepburn, and Henriksson. $E_c$ is the cohesive energy (eV/atom), $a_0$ is the lattice constant (Å), B is the bulk modulus (GPa), and $\Omega_0$ is the atomic volume ($\text{Å}^3/\text{atom}$).\\}
    \label{tab:2}
    \begin{tabular}{cccccc}
         \hline
         Property & DFT & Liyanage & B.J. Lee & Henriksson & Hepburn\\
         \hline
         $E_c$ & -5.82 & -5.82 & -4.8 & -5.46 & -5.11\\
         $a_0$ & 3.84 & 3.84 & 4.15 & 3.79 & 3.83\\
         B & 339.599 & 316.102 & 365.07 & 346.165 & 382.99\\
         $\Omega_0$ & 7.08 & 7.09 & 8.49 & 6.81 & 7.02\\
         \hline
    \end{tabular}
\end{table*}

\subsubsection{Bulk Fe (BCC) with C}

Effects of C on bulk structural properties are investigated through energy-volume curves at percentages ranging from 1 \% - 20 \%. This is the third test to evaluate the suitable potential. Carbon atoms are randomly distributed among the available octahedral sites of the BCC Fe structure. Octahedral interstitial sites with at least one lattice parameter apart were selected for possible C interstitials such that the interaction between C – C atom pairs would be minimized. To ensure adequate availability of octahedral sites agreeing to these conditions  3$\times$3$\times$3 simulation cells are used. In this structure for 1 \% C contained 55 total numbers of atoms with 1 C atom. For 5 \% C 57 atoms with 3 C atoms. Likewise,  for 8 \% 58 atoms with 4 C atoms, 10 \% 59 atoms with 5 C, 12 \% 60 atoms with 6 C, 15 \% 62 atoms with 8 C and for 20 \% 65 atoms with 11 C atoms are included.

\begin{figure}[H]
    \centering
    % First subfigure
    \begin{subfigure}[b]{0.23\textwidth} % Allocate 30% of the text width
        \centering
        \includegraphics[width=\textwidth]{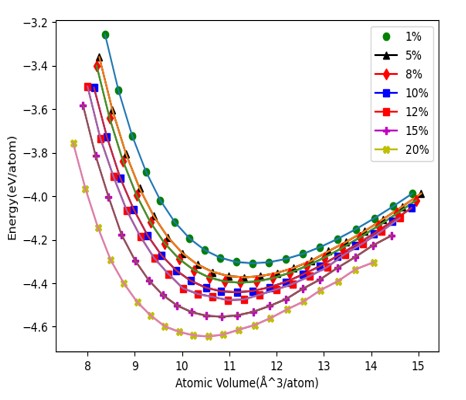}
        \caption{}
        \label{fig:subfig1}
    \end{subfigure}
    % Second subfigure
    \begin{subfigure}[b]{0.23\textwidth}
        \centering
        \includegraphics[width=\textwidth]{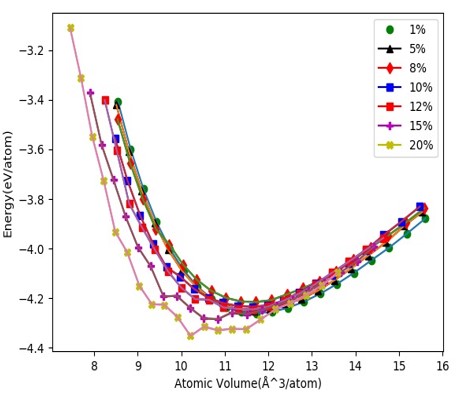}
        \caption{}
        \label{fig:subfig2}
    \end{subfigure}
    % Third subfigure
    \begin{subfigure}[b]{0.23\textwidth}
        \centering
        \includegraphics[width=\textwidth]{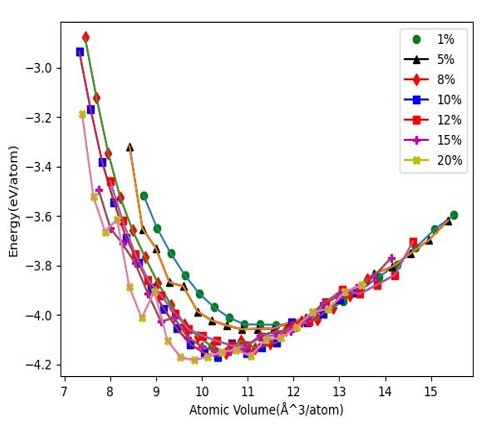}
        \caption{}
        \label{fig:subfig3}
    \end{subfigure}
    %fourth figure
    \begin{subfigure}[b]{0.23\textwidth}
        \centering
        \includegraphics[width=\textwidth]{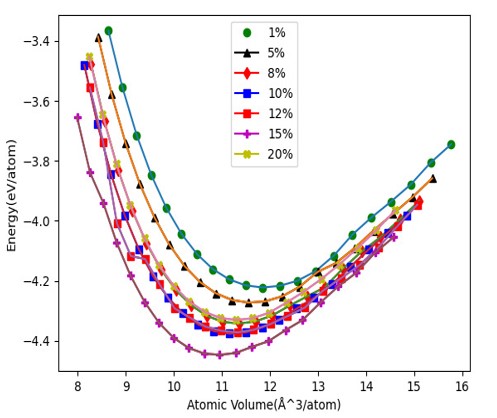}
        \caption{}
        \label{fig:subfig4}
    \end{subfigure}
    \caption{EV curves of BCC Fe-C (a) Liyanage (b) B.J. Lee (c) Hepburn (d) Henriksson varying C \%}
    \label{fig:EV curves of BCC Fe-C}
\end{figure}

\subsubsection{DFT curves for 10 \% and 20 \%}

\begin{figure}[!h]
    \centering
    % First subfigure
    \begin{subfigure}[b]{0.46\textwidth}
        \centering
        \includegraphics[width=\textwidth]{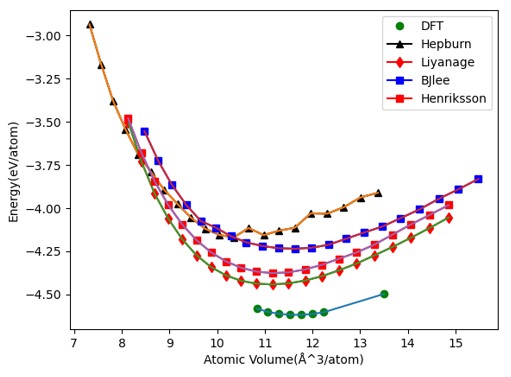}
        \caption{}
        \label{fig:subfig1}
    \end{subfigure}
    % Second subfigure
    \begin{subfigure}[b]{0.46\textwidth}
        \centering
        \includegraphics[width=\textwidth]{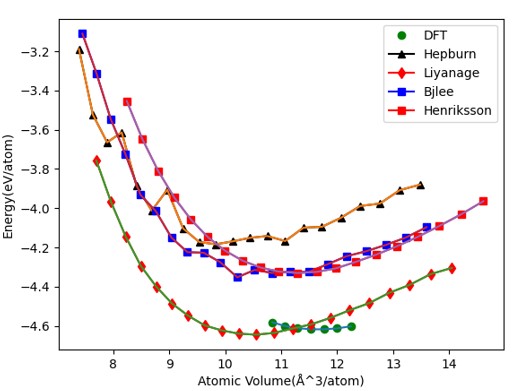}
        \caption{}
        \label{fig:subfig2}
    \end{subfigure}
    \caption{BCC Fe with (a) 10  \% C impurities (b) 20 \% C impurities}
    \label{fig:BCC Fe with 10 C impurities 20 C impurities}
\end{figure}

DFT energy - volume curves for C percentages 10 \%and 20 \% are obtained and compared with the curves of the interatomic potentials. Structural properties including lattice constant and bulk modulus for bulk Fe as a function of octahedral C concentration Table \ref{tab:4}, which parameters are extracted from those EV curves given in Figure \ref{fig:EV curves of BCC Fe-C}. The EV curves of 10 \% and 20 \% C concentrations are given in Figure \ref{fig:BCC Fe with 10 C impurities 20 C impurities}.

For 0 \% C, B.J. Lee predicts good agreement with experimental value. When increasing the C composition Henriksson and Liyanage interatomic potentials exhibit good results for 1 \% and 5 \% respectively. The values of bulk modulus from Henriksson interatomic potential are comparable with those reported in previous studies using MD simulation (179 GPa, 214 GPa for 1 \% and 5 \% respectively)\cite{henriksson2013atomistic}. For 10 \% C Liyanage and Henriksson potentials predicted bulk moduli which are similar to the DFT values. There is an approximately less than 10 \% change in the bulk modulus over the range of C percentages. Between 10 \% C and 15 \% C the bulk modulus increases in the Liyanage potential. Also, between 12 \% C and 20 \% C the bulk modulus increases in Henriksson's potential. However, in the Liyanage interatomic potential, the bulk modulus fluctuates when C percentage is increased. 

To validate the suitable potential for the Fe-C system four tests were done so far. According to the EV curves and structural properties a final interatomic potential was selected from four potentials. From the EV curves of 10 \% and 20 \% C, it is evident that the Liyanage interatomic potential is closer to the DFT values as well as it has a smooth EV curve. The smoothness of the EV curve is an indication of the robustness of the interatomic potential. Therefore, the Liyanage potential was selected to investigate the properties of steel NW.

\begin{table*}[h]
    \centering
    \caption{Bulk modulus B and the Lattice Constant $a_0$ for Bulk Iron (Fe) as a function of octahedral C content is listed. Bulk modulus (GPa) and Lattice constant (Å) of all the C percentages for each potential obtained. All C concentrations are in atomic percent (at.\%)\\}\label{tab:3}
    \resizebox{\textwidth}{!}{%
    \begin{tabular}{ccccccccccc}
        \hline
        C\% & \multicolumn{2}{c}{1\%} & \multicolumn{2}{c}{5\%} & \multicolumn{2}{c}{8\%} & \multicolumn{2}{c}{12\%} & \multicolumn{2}{c}{15\%} \\
        \cline{2-11}
         & B & $a^\circ$ & B & $a^\circ$ & B & $a^\circ$ & B & $a^\circ$ & B & $a^\circ$ \\
        \hline
        B.J. Lee & 162.5 & 2.88 & 166.4 & 2.91 & 151.76 & 2.93 & 155.45 & 2.94 & 149.28 & 2.92 \\
        Liyanage & 178.5 & 2.86 & 180.01 & 2.88 & 177.98 & 2.89 & 180.32 & 2.9 & 184.63 & 2.92 \\
        Henriksson & 186.8 & 2.89 & 176.06 & 2.9 & 176.21 & 2.9 & 172.01 & 2.93 & 174.76 & 2.93 \\
        Hepburn & 164.54 & 2.87 & 171.23 & 2.84 & 185.88 & 2.828 & 154.02 & 2.87 & 155.53 & 2.9 \\
        \hline
    \end{tabular}
    }
    \label{tab:properties}
\end{table*}

\begin{table*}[h]
    \centering
    \caption{Bulk modulus B and the Lattice Constant $a_0$ for Bulk Iron (Fe) for 10\% C and 20\% C is listed. Simulated DFT values given in parentheses. Bulk modulus (GPa) and Lattice constant (Å) of all the C percentages for each potential obtained.\\}\label{tab:4}
    \begin{tabular}{ccccc}
         \hline
         C\% & \multicolumn{2}{c}{10\%} & \multicolumn{2}{c}{20\%}\\
         \cline{2-5}
        & B (179.99) & $a\circ$ & B (182.98) & $a\circ$\\
         \hline
         B.J. Lee & 155.66 & 2.94 & 168.19 & 2.9\\
         Liyanage & 178.25 & 2.9 & 173.13 & 2.94\\
         Henriksson & 178.19 & 2.9 & 186.89 & 2.9\\
         Hepburn & 182.4 & 2.828 & 169.17 & 2.87\\
         \hline
    \end{tabular}
\end{table*}

\subsection{Steel nanowire simulations}

\subsubsection{Uni-axial tensile test}

To clearly show the effect of temperatures on tensile behaviours of steel nanowires, tensile processes at different temperatures were simulated. Here the temperatures of 0.1 K, 300 K, 600 K and 900 K were carried out.

\begin{figure}[H]
    \centering
    % First subfigure
    \begin{subfigure}[b]{0.46\textwidth} 
        \centering
        \includegraphics[width=\textwidth]{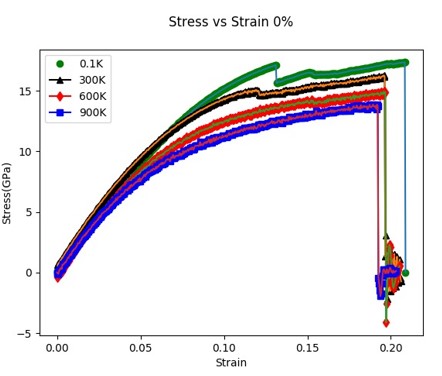}
        \caption{}
        \label{fig:subfig1}
    \end{subfigure}
    % Second subfigure
    \begin{subfigure}[b]{0.46\textwidth}
        \centering
        \includegraphics[width=\textwidth]{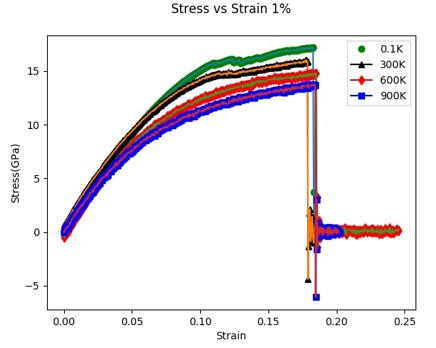}
        \caption{}
        \label{fig:subfig2}
    \end{subfigure}
    % Third subfigure
    \begin{subfigure}[b]{0.46\textwidth}
        \centering
        \includegraphics[width=\textwidth]{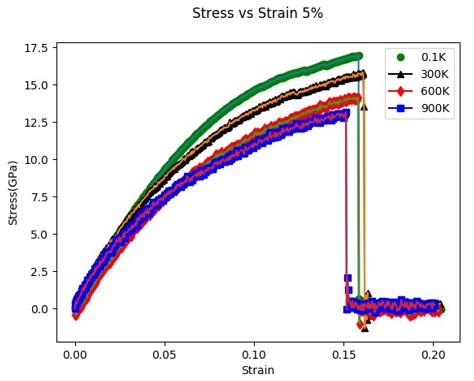}
        \caption{}
        \label{fig:subfig3}
    \end{subfigure}
    %fourth figure
    \begin{subfigure}[b]{0.46\textwidth}
        \centering
        \includegraphics[width=\textwidth]{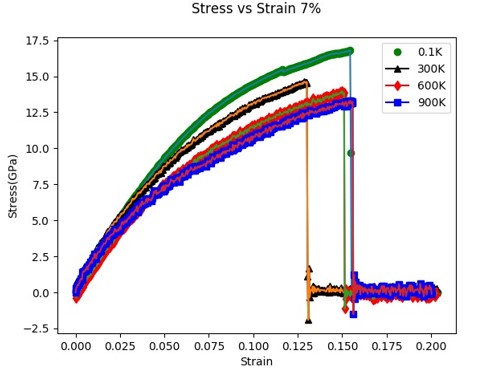}
        \caption{}
        \label{fig:subfig4}
    \end{subfigure}
    %fifth figure
    \begin{subfigure}[b]{0.46\textwidth}
        \centering
        \includegraphics[width=\textwidth]{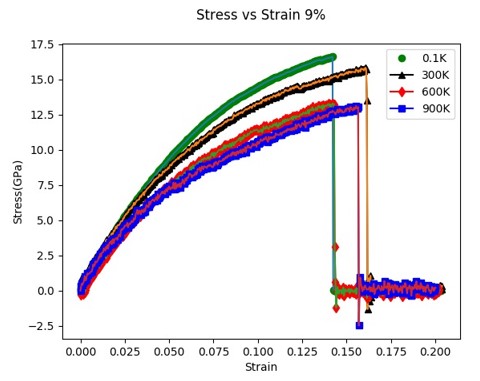}
        \caption{}
        \label{fig:subfig4}
    \end{subfigure}
    %sixth figure
    \begin{subfigure}[b]{0.46\textwidth}
        \centering
        \includegraphics[width=\textwidth]{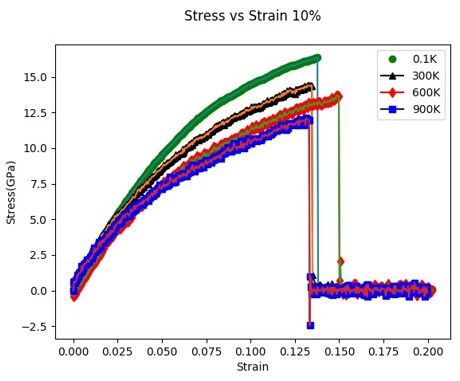}
        \caption{}
        \label{fig:subfig4}
    \end{subfigure}
    \caption{Stress-strain behaviour of (a) 0\% C (b) 1\% C (c) 5\% C (d) 7\% C (e) 9\% C (f) 10\% C with varying the temperature.}
    \label{fig:Stress-strain behaviour of C percentages}
\end{figure}

\begin{table}[H]
    \caption{Variation of (a) Maximum Stress (b) Young's modulus (c) Yield strength with C percentage at different temperatures.\\}\label{tab:5}
    \centering
        \begin{subtable}[b]{\textwidth}
            \caption{}
            \label{tab:5a}    
                \begin{tabular}{ccccccc}\hline
                    Temperature & 0\% & 1\% & 5\% & 7\% & 9\% & 10\%\\\hline
                    0.1 & 17.34878 & 17.17402 & 16.94185 & 16.79105 & 16.60395 & 16.35261\\
                    300 & 16.26874 & 16.00461 & 15.82892 & 14.62847 & 14.83893 & 14.46696\\
                    600 & 14.94656 & 14.78004 & 14.15937 & 13.9785 & 13.2798 & 13.69757\\
                    900 & 13.87173 & 13.76914 & 13.15636 & 13.2917 & 13.11466 & 12.09726\\\hline
              \end{tabular}
        \end{subtable}
        \begin{subtable}[b]{\textwidth}
            \caption{}
            \label{tab:5b}
                \begin{tabular}{ccccccc}\hline
                    Temperature & 0\% & 1\% & 5\% & 7\% & 9\% & 10\%\\\hline
                    0.1 & 209.2107 & 215.2538 & 226.4835 & 227.2259 & 232.0903 & 232.89\\
                    300 & 210.8864 & 212.9165 & 198.7577 & 207.5669 & 198.7664 & 222.8033\\
                    600 & 211.1754 & 206.6264 & 196.5188 & 196.1449 & 197.1035 & 196.4025\\
                    900 & 195.8466 & 194.812 & 204.967 & 196.4865 & 196.3095 & 186.1647\\\hline
                \end{tabular}
        \end{subtable}
        \begin{subtable}[b]{\textwidth}
            \caption{}
            \label{tab:5c}
                \begin{tabular}{ccccccc}\hline
                    Temperature & 0\% & 1\% & 5\% & 7\% & 9\% & 10\%\\\hline
                    0.1 & 10.27272 & 7.178535 & 6.74986 & 6.408371 & 6.387697 & 6.284072\\
                    300 & 10.48959 & 8.419011 & 7.324384 & 7.097678 & 6.911509 & 6.504372\\
                    600 & 7.796948 & 5.829023 & 5.373873 & 5.06595 & 4.767854 & 4.353248\\
                    900 & 5.571099 & 5.065464 & 4.649311 & 4.208726 & 3.857505 & 5.222194\\\hline
            \end{tabular}
        \end{subtable}
\end{table}

Figure \ref{fig:Stress-strain behaviour of C percentages} demonstrates the stress-strain curves during the tensile process at different temperatures, ranging from 0.1 to 900 K, in the [001] loading direction. The strain rate of $1 \times 10^8 \, \mathrm{s}^{-1}$ was applied for all temperature cases. It can be seen from Table~\ref{tab:5a} that, in general, the maximum stress decreases as the temperature increases for all C percentages. Tensile stress increases up to a critical value and then suddenly drops as the strain continues to increase. At lower temperatures, the stress-strain relationship behaves linearly. However, at higher temperatures ($> 300 \, \mathrm{K}$), the stress-strain curves exhibit nonlinear behavior. This is attributed to the increased atomic vibrations at elevated temperatures, which cause Fe-Fe and Fe-C bonds to deform more readily. During the stretching of the nanowire, the response of stress to strain presents a nonlinear performance due to this thermal-induced behavior. Such nonlinearity in stress-strain responses has also been observed in other metallic nanowires~\cite{li2017molecular}.

\begin{figure}[H]
    \centering
    \begin{subfigure}[b]{0.18\textwidth}
        \centering
        \includegraphics[width=\textwidth]{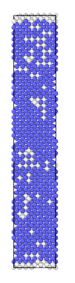}
        \caption{}
        \label{fig:subfig1}
    \end{subfigure}
    \hspace{0.02\textwidth}
    \begin{subfigure}[b]{0.18\textwidth}
        \centering
        \includegraphics[width=\textwidth]{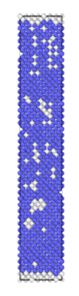}
        \caption{}
        \label{fig:subfig2}
    \end{subfigure}
    \hspace{0.02\textwidth}
    \begin{subfigure}[b]{0.18\textwidth}
        \centering
        \includegraphics[width=\textwidth]{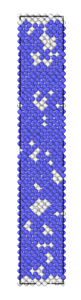}
        \caption{}
        \label{fig:subfig3}
    \end{subfigure}
    \hspace{0.02\textwidth}
    \begin{subfigure}[b]{0.18\textwidth}
        \centering
        \includegraphics[width=\textwidth]{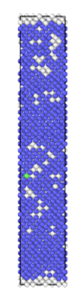}
        \caption{}
        \label{fig:subfig4}
    \end{subfigure}
    \caption{At 5\% C the equilibrated samples (a) 0.1 K, (b) 300 K, (c) 600 K, and (d) 900 K.}
    \label{fig:5 perc C the equilibrated samples}
\end{figure}

Common Neighbour Analysis (CNA) is used to clarify the local environment of the atoms in a crystal structure. Current study observed that when increasing the temperature, ultimate tensile strength is decreased. When visualized the structural changes and the defect formations in the nanowire could be easily seen with CNA. It identifies the local structure of the atoms by considering the number of neighbours.The initial atomic configuration of  5 \% C at 0.1 to 900 K are shown in the Figure \ref{fig:5 perc C the equilibrated samples}. Before the deformation process, defects are formed at each temperature. Blue represents the BCC phase and white represents unknown structure. With increased temperature the number of atoms in unknown structure increases. It reduces the strength of the steel NW which in turn fractures earlier than its pure counterpart.

\begin{figure}[H]
    \centering
    \begin{subfigure}[b]{0.225\textwidth}
        \centering
        \includegraphics[width=\textwidth]{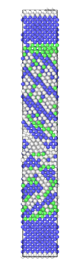}
        \caption{}
        \label{fig:subfig1}
    \end{subfigure}
    \hspace{0.02\textwidth}
    \begin{subfigure}[b]{0.22\textwidth}
        \centering
        \includegraphics[width=\textwidth]{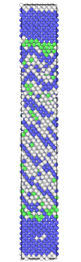}
        \caption{}
        \label{fig:subfig2}
    \end{subfigure}
    \hspace{0.02\textwidth}
    \begin{subfigure}[b]{0.21\textwidth}
        \centering
        \includegraphics[width=\textwidth]{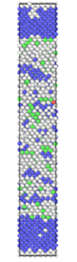}
        \caption{}
        \label{fig:subfig3}
    \end{subfigure}
    \hspace{0.02\textwidth}
    \begin{subfigure}[b]{0.215\textwidth}
        \centering
        \includegraphics[width=\textwidth]{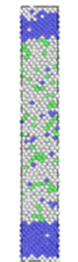}
        \caption{}
        \label{fig:subfig4}
    \end{subfigure}
    \caption{Atomic behaviour of 1\% C at the Ultimate Tensile Strength region (a) 0.1 K (b) 300 K (c) 600 K and (d) 900 K.}
    \label{fig:Atomic behaviour of 1 perc C}
\end{figure}

The growth of the dislocations at 1\% C respective to the increase of temperature shown in the Figure \ref{fig:Atomic behaviour of 1 perc C}. Amount of slip planes increases as the temperature increases. The propagation of slip planes occurs respectively when NW reaches uniaxial tensile strength Figure \ref{fig:Atomic behaviour of 1 perc C}. Therefore, a gradual BCC to FCC phase transition can be observed at Figure \ref{fig:Atomic behaviour of 1 perc C}. At 900 K atomic behavior drastically changes compared to 0.1 K and shows 60.7 \% of unknown structures. The BCC and FCC structures are present in percentages of 28.5 \% and 10.8 \% respectively. This results in decreasing the uniaxial tensile strength, with the increase of temperature at a constant C percentage.

The evolution of Young’s modulus and yield strength of steel nanowires with respect to temperature is obtained from the tensile stress–strain curves.  Here yield strength is calculated at the intersection point making a 0.2 \% offset to the linear region and Young’s modulus is determined by the slope of the elastic region of tensile stress–strain curves with the strain $<$1.2 \% using linear regression \cite{li2017molecular}. Table \ref{tab:mechanical property comparison} presents variation of the Young’s modulus and the yield strength extracted from the curves as a function of temperature at different C percentages.

\begin{figure}[!h]
    \centering
    \begin{subfigure}[b]{0.21\textwidth}
        \centering
        \includegraphics[width=\textwidth]{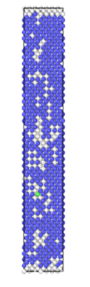}
        \caption{}
        \label{fig:subfig1}
    \end{subfigure}
    \hspace{0.02\textwidth}
    \begin{subfigure}[b]{0.22\textwidth}
        \centering
        \includegraphics[width=\textwidth]{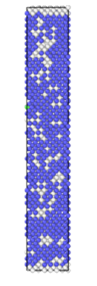}
        \caption{}
        \label{fig:subfig2}
    \end{subfigure}
    \hspace{0.02\textwidth}
    \begin{subfigure}[b]{0.24\textwidth}
        \centering
        \includegraphics[width=\textwidth]{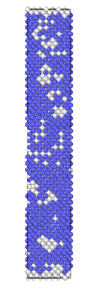}
        \caption{}
        \label{fig:subfig3}
    \end{subfigure}
    \hspace{0.02\textwidth}
    \begin{subfigure}[b]{0.215\textwidth}
        \centering
        \includegraphics[width=\textwidth]{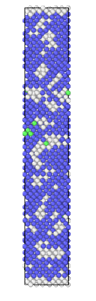}
        \caption{}
        \label{fig:subfig4}
    \end{subfigure}
    \caption{Atomic behavior at yield point, corresponds to increase of temperature at 10\% C (a) 0.1 K (b) 300 K (c) 600 K and (d) 900 K}
    \label{fig:Atomic behaviour at 10 perc C}
\end{figure}

It is clearly seen from table \ref{tab:mechanical property comparison} that both Young’s modulus and yield strength decrease with increasing temperature. Young’s modulus increases in the range of 0.1 – 300 K while decreases at 600 - 900 K at all the C percentages. Young’s modulus of the NW shows a different behaviour at $>$600 K. Nw comes to its maximum Young’s modulus at 300 K. After 600 K Young’s modulus no longer increases. It can be identified that the NW has a limit of the directly proportional relationship of Young’s modulus vs C percentage 600 K is a temperature beyond the limit. When increasing the temperature, the atoms gain energy. Then atoms become energetic, which is sufficient to break the bonds. This causes the decrease in the Young’s modulus. It is precise that steel now has more stiffness property when increasing the C \% from 0 - 10 at the temperature range from 0.1 K – 300 K. At higher temperatures steel NW does not predict high stiffness, even though the C \% is increased.
Figure \ref{fig:Atomic behaviour at 10 perc C} shows the atomic behavior at the yield point, corresponding to an increase of temperature at 10 \% C. As the temperature is increased, a greater number of atoms gain enough energy to overcome the energy barrier. At the yield point, the growth of deformation propagates  and permanent deformation takes place. Even though these Steel NWs have more stiffness, dislocations propagate rapidly. Therefore, NW reaches the plastic region quicker. This causes a decrease in mechanical properties and stress for twining propagation. This phenomenon can be observed in each C percentage given in the table \ref{tab:mechanical property comparison}
The values of BCC Fe of yield strength comparable with those reported in previous studies using MD simulations (at 300 K $\approx$10 GPa, at 600 K $\approx$5 GPa, at 900 K $\approx$4 GPa) \cite{li2017molecular}, at 0.1 K ab initio calculations (12.7 GPa)\cite{li2017molecular}. Young’s modulus also compared with ab initio calculation at 0.1 K (155 GPa) \cite{li2017molecular}. 

\begin{figure}[H]
    \centering
    \begin{subfigure}[b]{0.14\textwidth}
        \centering
        \includegraphics[width=\textwidth]{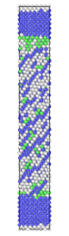}
        \caption{}
        \label{fig:subfig1}
    \end{subfigure}
    \hspace{0.01\textwidth}
    \begin{subfigure}[b]{0.156\textwidth}
        \centering
        \includegraphics[width=\textwidth]{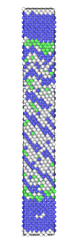}
        \caption{}
        \label{fig:subfig2}
    \end{subfigure}
    \hspace{0.01\textwidth}
    \begin{subfigure}[b]{0.143\textwidth}
        \centering
        \includegraphics[width=\textwidth]{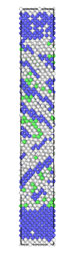}
        \caption{}
        \label{fig:subfig3}
    \end{subfigure}
    \hspace{0.01\textwidth}
    \begin{subfigure}[b]{0.147\textwidth}
        \centering
        \includegraphics[width=\textwidth]{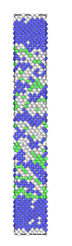}
        \caption{}
        \label{fig:subfig4}
    \end{subfigure}
    \hspace{0.01\textwidth}
    \begin{subfigure}[b]{0.145\textwidth}
        \centering
        \includegraphics[width=\textwidth]{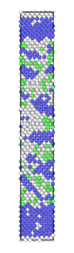}
        \caption{}
        \label{fig:subfig5}
    \end{subfigure}
    \hspace{0.01\textwidth}
    \begin{subfigure}[b]{0.153\textwidth}
        \centering
        \includegraphics[width=\textwidth]{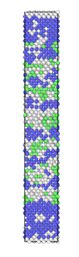}
        \caption{}
        \label{fig:subfig6}
    \end{subfigure}
    
    \caption{Atomic behavior at the Ultimate Tensile Strength region, corresponding to increasing C percentage at a constant temperature of 300 K: (a) 0\% C, (b) 1\% C, (c) 5\% C, (d) 7\% C, (e) 9\% C, and (f) 10\% C.}
    \label{fig:Constant temperature Atomic behaviour}
\end{figure}

The evolution of the structure with the increase of C percentage is shown in Figure \ref{fig:Constant temperature Atomic behaviour}. When increasing the C percentage, the uniaxial tensile strength decreases. This happens due to the slip plane propagation as the increase of C. Slip planes become thicker when increasing the C \%. As known the interaction between stress fields formed by the slip dislocations interferes with the dislocation motion by repulsive and attractive interactions . 
With the growth of high dislocations NW’s reach to its maximum stress and abruptly drops the stress.  This phenomenon causes the uniaxial tensile strength to decrease. This can be observed in each C percentage as illustrated in Figure \ref{fig:Stress-strain behaviour of C percentages}. 

Adding C atoms to the existing Fe structure, significantly affects the spacing between atoms. (0 C\% - 10 C\%). When increasing C Percentage Young’s modulus increases. This happens as a result of decrease of atomic bond space, since lower bond distances have higher strength. Therefore, steel NW gets more stiffness when increasing C\% in the range of 0.1 - 300 K.

\begin{table*}[h]
    \centering
    \caption{Comparison of mechanical properties of steel nanowires with bulk mechanical properties.}
    \begin{tabular}{ccccc}
         \hline
         Mechanical Property & \multicolumn{2}{c}{Bulk Steel (Experimental)} & \multicolumn{2}{c}{Current Study} \\
         \cline{2-5}
         & 1\% & 5\% & 1\% & 5\% \\
         \hline
         Yield Stress & 220 MPa & 455 MPa & 8.42 GPa & 7.32 GPa \\
         Young's Modulus & 400 MPa & 825 MPa & 212.91 GPa & 198.76 GPa \\
         \hline
    \end{tabular}
    \label{tab:mechanical property comparison}
\end{table*}

Table \ref{tab:mechanical property comparison} presents the compared values of steel nanowires of yield strength and young’s modulus with those bulk steel reported in previous studies using experimental values \cite{bramfitt1998structure}. These values might be changed because the experiment was done using a cylindrical wire. However we can clearly see nano steel  has greater yield and Young's modulus compared to the steel bulk. When increasing the C\% both properties of steel nanowire decreased but these values are greater than the bulk steel properties.

%% file: Sections/Conclusion.tex
In the present work we have evaluated the four interatomic potential models to describe the mechanical properties of steel nanowires. The structural property predictions of BCC bulk Fe, hypothetical $B_1$ rock salt structure and BCC iron with varying percentages of C atoms by the potentials were used to select the most suitable interatomic potential. Among the four potentials, MEAM potential by Liyanage et al.  gave the best predictions which were in good agreement with experimental/first principles calculated values. Therefore, the potential by Liyanage et al. \cite{Liyanage2014_Cementite}. was chosen for the simulation of steel nanowires. The uniaxial tensile test of steel nanowires was simulated using molecular dynamics at 0.1 K, 300 K, 600 K and 900 K with C percentages of 0\%, 1\%, 5\%, 7\%, 9\% and 10\%. The Young’s modulus increased with the increase of C percentages up to 300 K and decreased from 600 - 900 K. However, the yield point decreased with an increase of C percentage at all the temperatures. Common Neighbour Analysis (CNA) was used to identify the microstructural changes in the MD simulations. Through CNA we found that the rapid formation of slip planes is assisted by the C atoms that were placed in the interstitial positions. CNA also showed that with the increase of temperature the rapid structural deterioration of the Fe BCC structure with increased strain. Also compared to the bulk steel, steel nanowire predicted greater mechanical properties. The present work will be extended up to further simulations of the steel NW s. The dependence of the size and strain rates against the NW properties are to be investigated in the future work.